# Directional self-propelled transport of coalesced droplets on a superhydrophilic cylindrical wire

**Leyun Feng[1,3], Youhua Jiang[1,3], Christian Machado[1], Wonjae Choi[2], Neelesh A. Patankar[1], Kyoo-Chul Park[1*]**

[1] Department of Mechanical Engineering, Northwestern University, Evanston, IL 60208, USA

[2] Gildart Haase School of Computer Sciences and Engineering, Fairleigh Dickinson University, Teaneck, NJ 07666, USA

[3] These authors contributed equally: Leyun Feng, Youhua Jiang.

Droplets coalescing on flat surfaces tend to end up with the smaller droplet migrating into the larger one. We report a counter-intuitive droplet coalescence pattern on a superhydrophilic cylindrical wire, where the larger droplet is pulled toward the smaller one. Consequently, the center of the combined mass significantly moves toward, often beyond, the original location of the smaller droplet. This phenomenon occurs primarily because the viscous friction that a droplet experiences on a cylindrical wire is not positively correlated with the size of the droplet, unlike the droplets coalescing on flat surfaces. We conducted a dimensional analysis based on a mass-spring-damper (MSD) model. Our model predicts a variety of coalescence patterns as a function of the ratio between droplet sizes, and the prediction matches the experimental observation.

---

[*] To whom correspondence should be addressed. E-mail: kpark@northwestern.edu

# 1. Introduction

Liquid transport on cylindrical surfaces is important to various potential applications, such as fog collection, mist elimination, and filtration [1-4]. The devices used in these common processes are often mesh structured, composed of hundreds of single wires or fibers. These meshes can filter a mixture of various phases into separate ones by allowing one phase to pass through the mesh while capturing the other. Clogging that occurs between individual wires, however, would influence the mesh openness and reduce the separation efficiency.

Passive removal of clogging droplets may be possible when such droplets spontaneously migrate from the clogging area toward the periphery of the mesh so that the mesh can continue to operate. There have been multiple reports on the spontaneous migration of droplets on surfaces. Papers by Lorenceau & Quéré or Li & Thoroddsen focused on the curvature-driven migration of droplets on conic wires [5,6]. Such mechanism is utilized by natural plants such as cactus, whose unique conical-based hierarchical structure collects water from fog and route the droplets toward the base [7]. The jumping droplet phenomenon is another mechanism which allows surfaces to remove water droplets from superhydrophobic surfaces without active energy input [8]. One can imbue surfaces with a gradient of surface energy [10], surface charge density [9], or temperature [11]. Droplets migration based on active control such as light [12] or magnetic field [13] has also been a topic of research.

While all these reports are valuable discoveries, their applicability to mesh surfaces is somewhat limited. Conical wires need to be short for there to be significant gradient of curvature along the wires, making it impractical to fabricate large meshes of long conic wires. Superhydrophobic surfaces let droplets jump off only if the droplets do not coalesce between the surface texture leading to Wenzel state [14]. Rendering surfaces to have gradients of energy, charge, or temperature has the same limitation of conic wires. Active control of droplets has its own applications, but it is not applicable for large scale mesh surfaces that are supposed to capture fog droplets in deserts.

Further, the behavior of liquid on curved surfaces is often different from that on flat surfaces. While liquid on flat superhydrophilic surfaces simply spread to form a film, liquid on wires form the shape of a droplet even when the wires are superhydrophilic [15,16]. These droplets display interesting dynamics including the noted curvature-driven migration and wire-droplet elastocapillary interaction [17,18]. These studies show that the speed or even direction of motion of droplets can be drastically different on wires than on flat surfaces. Such difference was explained using the asymmetric effect of surface tension or elasticity of wires.

This paper reports a fast, directional, and self-propelled droplet transport without any geometry or chemical control. Such transport is based on less explored behavior of droplets, driven by the unique trait of droplets on cylindrical wires. Droplets deposited on surfaces form a contact line with the surface, and the resistance against motion is strong near the contact line. For self-similar droplets on a flat surface, the length of the contact line is proportional to the diameter of the droplet, making the droplet more resistant against motion as the droplet becomes larger. On the other hand, droplets on cylindrical wires form contact lines of the same length regardless of their size, as long as the droplet is larger than the wire. While the concept of a contact line becomes invalid on superhydrophilic wires, the size of the highly dissipative wedge zone of the droplet is still largely independent of the size of the droplet in such cases. Thus the resistance against motion is not strongly correlated with the size of the droplet. Such difference can cause a counter-intuitive motion that we report in this paper.

## 2. Coalescence on cylindrical wires

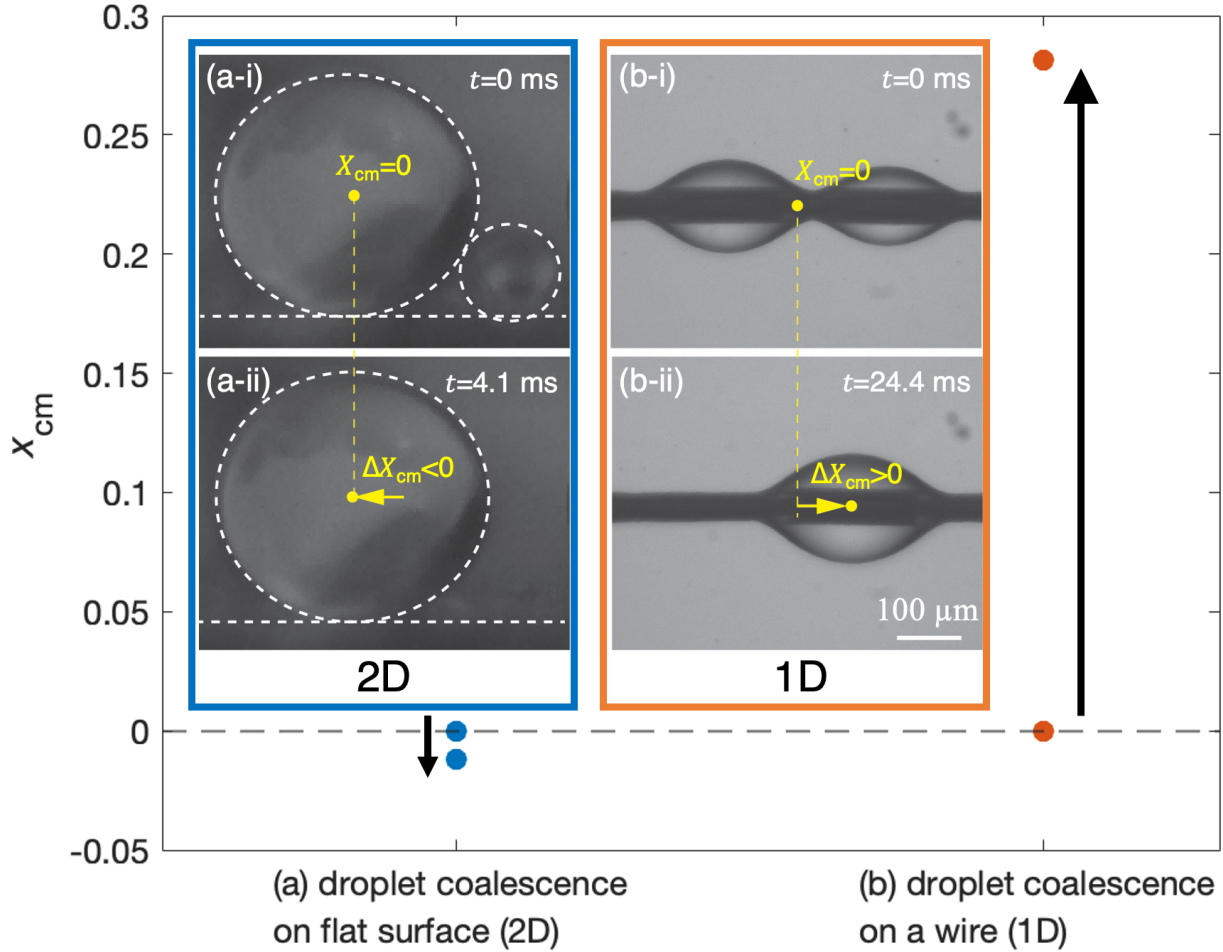

**Figure 1**. Coalescence on a flat surface and a cylindrical wire. (a) Droplets of DI water coaslescing on a superhydrophobic surface. (b) Droplets of DI water coalescing on a superhydrophilic cylindrical wire. Wire diameter is 127 μm. (c) Center of combined mass before and after coalescence. Inset shows the shape of the droplets before and after the event.

A simple experiment can demonstrate the difference in coalescence patterns on a flat surface and on a cylindrical wire as shown in Figure 1. Droplets of DI water are deposited on a superhydrophobic surface (Fig. 1(a)) or on a superhydrophilic wire (Fig. 1(b)). Regarding droplets on wires, we limit our discussion on barrel shape droplets only. Also, the droplet mass is limited up to $10^{-3}$ g (typically around $10^{-4}$ g) in order to keep the shape of the droplet reasonably axisymmetric. Surfaces were prepared by boehmitizing aluminum plates or wires, a process which renders aluminum surfaces nanotextured and superhydrophilic. Superhydrophobicity was selectively imbued by a fluoroaliphatic phosphoric acid process. Parent droplets were deposited by a sprayer (Houseables), and we let them grow by humidifiers (Pure enrichment).

The height $h$ and length $L$ of each droplet before and after coalescence were determined with video analysis tool Tracker (free video analysis and modeling tool built on the Open Source Physics (OSP) Java framework). Using the analytical expressions given by Carroll, B. J [19], we obtained the droplet volume $V$. Droplet mass was simply computed from $m = \rho V$, where $\rho$ is the density of the liquid. The center of mass migration can also be tracked in Tracker.

Figure 1 shows the result of coalescence. The coalescence patterns of droplets on superhydrophobic surfaces is as expected; the smaller droplet appears to be pulled into the larger one. This observation is intuitive; the larger droplet has correspondingly larger inertia and also a longer contact line than the smaller droplet. Also, the Laplace pressure due to the curvature of the droplet drives the smaller droplet to migrate toward the larger one. Collectively, the smaller droplet is pulled toward the larger one while the larger one resists motion [20-23]. As a result, the center of the combined mass moves toward the larger one but only barely. The coalescence patterns of droplets on superhydrophilic wires, on the other hand, is the opposite. The larger droplet is pulled toward the smaller one, forming the merger droplet with its center of mass closer to the smaller droplet. This is counter-intuitive, opposite to a simple prediction based on inertia or Laplace pressure.

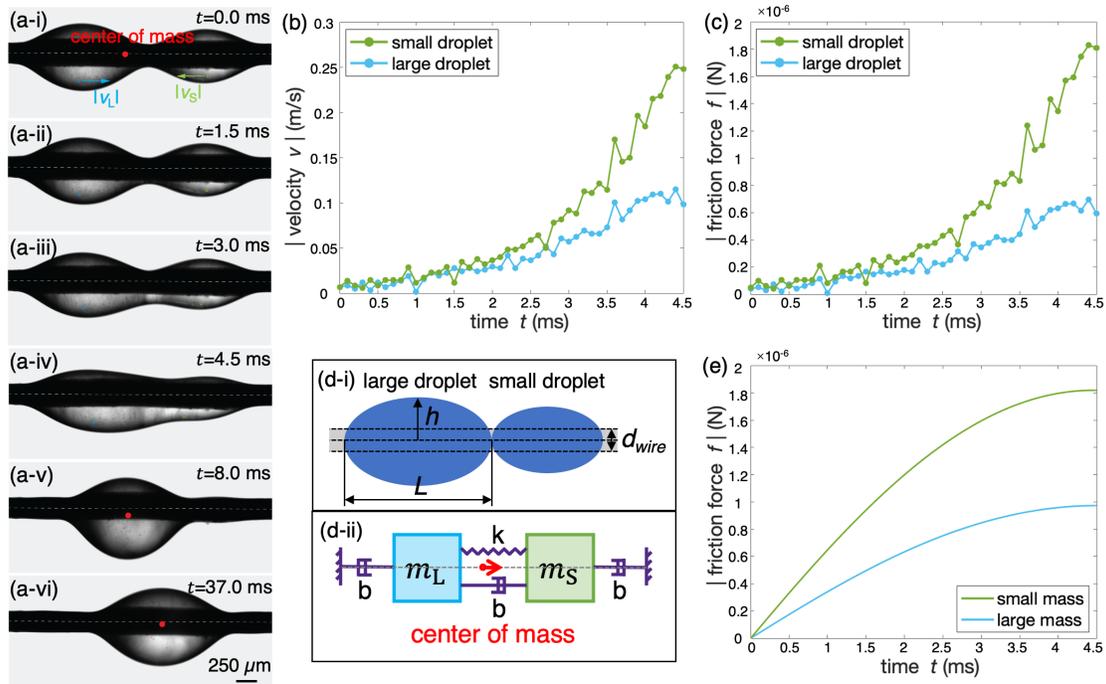

**Figure 2.** (a) Time lapse images of the coalescence event of DI water droplet containing micro-scale particles. Wire diameter is 250 μm. (b) Velocity of the two small droplets during the initial 4.5 ms of the event. (c) Friction that each droplet experiences during the same period, calculated by $F_{fric} = \mu \frac{v}{h} \pi d_{wire} L$. (d) Equivalence between droplet coalescence and a MSD model. (e) Friction force on the two droplets predicted by the MSD model.

Figure 2(a) shows the time-lapse images of the coalescence event described in Figure 1. During the early phase of the coalescence up to t = 11 ms, the smaller droplet almost disappears while the larger droplet grows noticeably. Coalescence event up to this moment agrees with the common model based on Laplace pressure, which predicts that the smaller droplet should be pulled into the larger one. What is surprising is that the merger droplet moves rightward after that, despite the obvious leftward flow from the smaller parent droplet. The center of the combined mass, red dot in the figure, also moves rightward. The migration of the center of mass is unexpected because the coalescence is driven by the surface tension, which is the force of mutual interaction between two droplets. By Newtonian action-reaction principle, the magnitude of the pulling forces of one droplet on the other should be identical, and these self-canceling two forces cannot create any net acceleration of the droplet.

We claim that the phenomenon is driven by the unique mechanism of viscous dissipation for droplets on cylindrical wires. Most of viscous dissipation for a droplet moving on solid surfaces occurs near the wedge of the droplet if the surface is superhydrophilic. The length of the contact line or of the outer edge of a droplet is directly proportional to the size of the droplet when the surface is flat. In contrast, the length of the contact line or the edge of a small droplet on a cylindrical wire is precisely identical to the one of a larger droplet, because it is merely the circumference of the wire. Therefore, the viscous dissipation for a barrel-shaped droplet sliding with a given velocity on a cylindrical wire becomes mostly independent of the droplet size.

Lorenceau and Quéré [24] analyzed that, when a droplet moves on a superhydrophilic wire that is pre-wetted, viscous friction resisting the motion can be approximated to be a sum of two different dissipation terms, one occurring in a nanoscopic Hoffman wedge and the other in the macroscopic wedge. We ignore the dissipation in the nanoscopic Hoffman wedge, because such a thin layer can exist only on an atomically smooth surface while our superhydrophilic wires have rough texture. Dissipation in the macroscopic wedge can be estimated by integrating the shear stress inside the wedge. Lorenceau and Quéré conducted the integration to predict the friction would scale as

$$F_{fric} \sim \frac{k\pi d_{wire}\mu v L}{h} \quad \text{(Eq. 1)}$$

where $v$ is the speed of the droplet and $k$ ($k \approx \ln r/\alpha$, $\alpha \sim 10^{-9}$ m) is a logarithmic factor to prevent the divergence in integration at the edge. We ignore $k$ for the rest of this work as it is essentially a constant, which does not affect the scaling relationship. $d_{wire}$ is the wire diameter. $L$ and $h$ are the axial length of the droplet and the radial thickness of the droplet measured from the surface of the wire, respectively. $\mu$ is the viscosity of the liquid. The numerator of the expression shows that the viscous friction would be proportional to the velocity of the droplet, which is intuitive. It also indicates that the friction would be roughly proportional to the diameter of the wire, not the size of the droplet. Size of the droplet affects the friction by changing the wedge shape, whose effect is reflected in the ratio between the length and thickness of the droplet. Its influence is, however, limited; the ratio remains in a narrow range unless the droplet is extremely small or large. The $L/h$ ratio is between 6 and 8 for all trials in this work.

Figure 2(b) shows the velocity of each droplet, measured by tracking the two particles in each droplet marked as cyan and green. The tracking is conducted only up to the initial 4.5 ms because the two droplets quickly become one, after which it is impossible to track the droplets separately. Clearly, the smaller droplet is moving much faster than the larger droplet during the initial phase. This is not surprising, considering that both the Laplace pressure and inertia causes higher acceleration for the smaller droplet. What is unique for droplets on cylindrical wires is the viscous friction. Equation 1 predicts that, as the friction that a droplet experiences is virtually independent of its size, it depends only on the speed of the droplet. Figure 1(c) shows the estimated friction that each droplet experiences during the initial 4.5 ms. We believe that the noted difference in friction is what drives the counter-intuitive motion of the merger droplet. While the Laplace pressure and inertia both lead the smaller droplet to migrate toward the larger one, those two forces are both internal forces between the two droplets and thus they cannot create any motion of the center of the combined mass. The only external force that is applied on the system of two droplets is the friction applied by the wire on the droplets. Noted difference in friction thus is the main driver for the merger droplet.

### 3. Mass-spring-damper (MSD) model

Here we derive a simple mass-spring-damper (MSD) model to reason about the transition from surface to kinetic energy during coalescence. Inset of Figure 2(d) shows the schematic drawing. Each droplet is modeled as a mass. Two masses are connected through a spring, which is a simplified version of the surface tension. The masses will thus be pulled or pushed by surface tension, and the motion is resisted by viscous friction that is modeled by dampers.

We determine the value of each parameter as the following. Mass of the large droplet $m_L$ and small droplet $m_S$ is simply determined based on their size. $m_L = \rho V_L$ and $m_S = \rho V_S$, where is $\rho$ the DI water density. Spring constant $k$ is determined by equating the potential energy of this hypothetical spring to the difference between the surface energy of the parent droplets and of the merger droplet. More precisely, elastic potential energy stored in the model spring is

$$E = \frac{1}{2} k x_{k,0}^2 \qquad (Eq.\ 2)$$

where $x_{k,0}$ is the initial elongation of the spring.

We can equate this energy to the difference in the potential energy levels before and after coalescence, estimated to be

$$E = \gamma[4\pi(R_L^2 + R_S^2 - R_{L+S}^2) - 2\pi d_{\text{wire}}(R_L + R_S - R_{L+S})] \qquad (Eq.\ 3)$$

By equating the two terms, we obtained $k = \frac{2\gamma[4\pi(R_L^2 + R_S^2 - R_{L+S}^2) - 2\pi d_{\text{wire}}(R_L + R_S - R_{L+S})]}{x_{k,0}^2}$, where the volume-equivalent radii of the droplets are defined as $R_L = \sqrt[3]{\frac{3V_L}{4\pi}}$, $R_S = \sqrt[3]{\frac{3V_S}{4\pi}}$, and $R_{L+S} = \sqrt[3]{\frac{3V_{L+S}}{4\pi}}$.

Damping coefficient $b$ is obtained in a similar manner, by equating the friction force in a model damper to the friction force acting on the droplet by the wire. The friction model in a MSD is $F_{fric} = bv$, and by equating it to the viscous friction estimated by Equation 1, we obtain

$$b = \mu \frac{\pi d_{\text{wire}} L}{h} \qquad (Eq.\ 4)$$

where $\frac{L}{h}$ is between 6 and 8 as previously mentioned. As a result, damping coefficient b is nearly identical between the two parent droplets despite the difference in sizes.

We use Euler's method to iteratively track the motion of the coalescing droplets. For each iteration in time, $i$, we calculate an acceleration $a_{L,i}$ for the large droplet and $a_{S,i}$ for the small droplet which is assigned to the next incremental time step, $dt$. These accelerations are calculated using Newton's 2nd law:

$$a_{L,i} = \frac{-2bv_{L,i} + kx_{k,i} + bv_{S,i}}{m_L} \tag{Eq. 5}$$

$$a_{S,i} = \frac{-2bv_{S,i} - kx_{k,i} + bv_{L,i}}{m_S} \tag{Eq. 6}$$

Using Euler's method, we can find the starting velocity of the next time iteration from kinematics:

$$v_{L,i+1} = v_{L,i} + a_{L,i} dt \tag{Eq. 7}$$
$$v_{S,i+1} = v_{S,i} + a_{S,i} dt \tag{Eq. 8}$$

We can also find the starting position of the next iteration:

$$x_{L,i+1} = x_{L,i} + v_{L,i} dt + \frac{1}{2} a_{L,i} dt^2 \tag{Eq. 9}$$
$$x_{S,i+1} = x_{S,i} + v_{S,i} dt + \frac{1}{2} a_{S,i} dt^2 \tag{Eq. 10}$$

For each time iteration, we can determine the center of mass position by $x_{CM,i} = \frac{m_L x_{L,i} + m_S x_{S,i}}{m_L + m_S}$.

Finally, we set the initial condition as the following. Initial elongation of the spring is set to be $x_{k,0} = 2(R_L+R_S)$, the total length of the two-droplet system. The equilibrium length of the spring is $l = 0$. Initial velocities of the two masses are $v_{L,0} = v_{S,0} = 0$, considering that the coalescence begins from rest. To set the center of mass of the two droplets $x_{cm,0} = 0$, we set the initial locations of the two parent droplets as below:

$$x_{L,0} = \frac{-(x_{k,0}+l)m_S}{m_L+m_S} = \frac{-2(R_L+R_S)m_S}{m_L+m_S} \tag{Eq. 11}$$

$$x_{S,0} = \frac{(x_{k,0}+l)m_L}{m_L+m_S} = \frac{2(R_L+R_S)m_L}{m_L+m_S} \tag{Eq. 12}$$

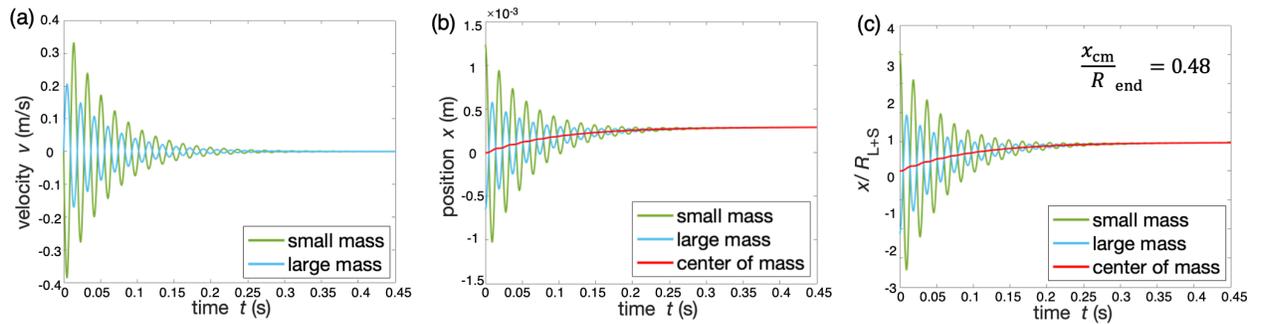

**Figure 3**. Prediction by the MSD model. (a) Velocity of two droplets from the initial phase until the steady state. Note each velocity is offset from zero, positive for the larger droplet and negative for the smaller droplet. (b) Migration of the two masses and also of the center of the combined mass. Masses in this model never merge into a single mass, but their locations gradually approach each other. (c) Dimensionless number $x/R_{L+S}$.

The friction force experienced by the two masses, predicted by this model, is in Figure 2D. The empirical and theoretical trends are similar, both in qualitative direction and quantitative magnitude. Figure 2D also clearly indicates that the viscous friction is significantly larger for the smaller droplet, due to its larger velocity. This difference should cause the two-mass system to move toward the smaller one. Figure 3 shows the long-term prediction from the model. Modeling the long-term coalescence process by distinct two masses may be a strong simplification, but the model captures the axial elongation-contraction cycles as the oscillation of the two point-masses successfully. The MSD model predicts that both masses display underdamped oscillations (Figure 3(a)), which matches the experimental observation. Velocity of the larger droplet is offset toward the positive side, meaning that the droplet displays net motion toward right. Velocity of the smaller droplet is offset toward the negative side, meaning that the droplet displays net motion toward left. While both velocity profiles display oscillatory patterns, velocity of the smaller droplet is larger in magnitude. As a result, the friction applied on the smaller droplet is higher than the one on the larger droplet, creating a net displacement toward the right direction (Figure 3(b)). Normalizing the displacement by the volume equivalent radius of the merger droplet gives the nondimensional displacement in Figure 3(c), which predicts that the migration of the merger droplet would be on the order of its body length, again matching the experimental observation.

## 4. Droplet upward movement

While the damping coefficient is largely independent of the droplet size, the viscous friction is. The pulling force from the surface tension, modeled as the spring force in the MSD, causes larger acceleration and velocity for the smaller droplet, making it experience larger viscous friction. As the difference in the friction is the driving force for the net migration of the droplet, we suspect there to be scaling relationship between nondimensional migration $x_{cm}/R_{L+S}$ and nondimensional ratio between droplet sizes $R_L/R_S$. Figure 4(e) shows the scatter plot of all experimental data we obtained on horizontal wires (solid green, blue, and yellow dots) on $x_{cm}/R_{L+S}$ - $R_L/R_S$ plane. The figure also shows the prediction by our MSD model (red hollow dots), which shows reasonable agreement with the experimental result.

When the Bond number ($Bo = \rho g r^2/\gamma$) of a droplet is small, the elastic potential energy outweighs the gravitational energy and thus upward movement can be achieved upon coalescence as shown in Figure 4(a-d). When the smaller droplet is placed above the larger one along the wire, the larger droplet still would be pulled toward the smaller one during coalescence, overcoming the gravitational force. Adding experimental data on inclined wires (black solid dots) in Fig. (e), all of them are below the predicted value (red hollow dots). This is because a part of energy transit to gravitational potential energy during the droplet coalescence.

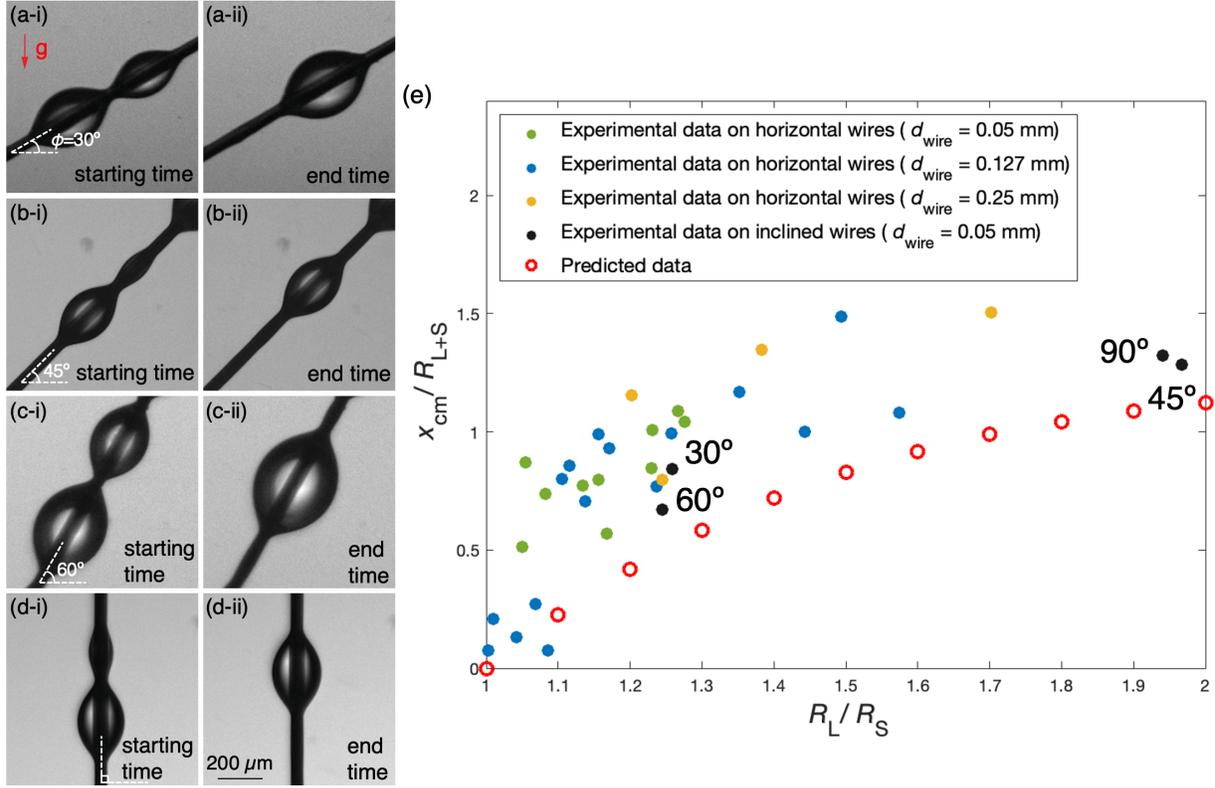

**Figure 4**. Regime map. (a-d) Upward movement of droplet overcoming gravitational force on wires with different inclined angles. (e) Regime map showing the relation between $x_{cm}/R_{L+S}$ and $R_L/R_S$. The uncertainly of $x_{cm}/R_{L+S}$ and $R_L/R_S$ for experimental data are within 5% and 10%, respectively. Solid dots shows the experimental data. Red hollow dots shows the data predicted by the MSD model.

## 5. Discussion and conclusion

In summary, we have reported a counter-intuitive droplet transport phenomenon along a superhydrophilic wire, where two droplets of distinct sizes merge to migrate toward or beyond the location of the smaller parent droplet. The center of the combined mass thus moves across substantial distance, on the order of the body length of the droplet itself. This phenomenon is caused primarily by the asymmetric resistance at the two end wedges. An MSD model is built to analyze the droplet movement and predict the relation between two important dimensionless numbers: $x_{cm}/R_{L+S}$ and $R_L/R_S$. Our finding may allow researchers and engineers to design better mechanisms for fog/dew collection, filtration, emulsion separation, or removal of toxic aerosols where the droplets interact with each other on cylindrical wires.

## 6. Acknowledgements

This work was partially supported by the Water Collaboration Seed Funds program of the Northwestern Center for Water Research.

The authors thank Dr. David Quéré, Dr. Gareth H. McKinley, and Dr. Jin-Tae Kim for their insightful comments.